\documentclass[aip,reprint]{revtex4-1}
\usepackage{array}
\usepackage{longtable}
\usepackage{multirow}
\usepackage{color}

\begin{document}


\title{Compilation of spectroscopic data of Radium (Ra I and Ra II)} 



\author{U. Dammalapati}
\email[]{dammalapati.umakanth.a8@tohoku.ac.jp}
\altaffiliation{Present address: Cyclotron and Radioisotope Center (CYRIC), Tohoku University, 6-3 Aramaki-aza Aoba, Aoba-ku, Miyagi 980-8578, Japan}

\author{K. Jungmann}
\author{L. Willmann}
\email[]{l.willmann@rug.nl} \affiliation{Van Swinderen Institute,
Faculty of Mathematics and Natural Sciences, University of
Groningen, Zernikelaan 25, 9747 AA Groningen, The Netherlands}


\date{\today}

\begin{abstract}
Energy levels, wavelengths, lifetimes and hyperfine structure
constants for the isotopes of the first and second spectra of
radium, Ra I and Ra II have been compiled. Wavelengths and wave
numbers are tabulated for $^{226}$Ra and for other Ra isotopes.
Isotope shifts and hyperfine structure constants of even and odd-A
isotopes of neutral radium atom and singly ionized radium are
included. Experimental lifetimes of the states for both neutral
and ionic Ra are also added, where available. The information is
beneficial for present and future experiments aimed at different
physics motivations using neutral Ra and singly ionized Ra.
\end{abstract}



\maketitle 

\section{Introduction}
Radium (Ra), a radioactive element was discovered by P. Curie, M. Curie and G. Bemont in the year 1898~\cite{Curie1898}. It was
extracted by chemical means from pitchblende or uraninite. About
one gram (1g) of radium is present in 7 tons of pitchblende. The
activity of 1g of $^{226}$Ra was defined later as the unit of
radioactivity, Curie, Ci (1~Ci~=~$3.7\times10^{10}$~disintegrations
per second (dis/s)). The term \emph{radioactivity} was introduced
by Mme. Marie Curie. Ritz in 1908 classified two
series of lines of neutral radium, Ra~I~\cite{Ritz1908}. In 1911,
radium was isolated by Mme. Curie and Debierne, who found it to be
a brilliant white solid~\cite{Lide2008}. Some energy levels and
line classifications of Ra~I were given by Fues in
1920~\cite{Fues1920}. In 1922, Fowler gave classifications of some
series of lines in both Ra I and singly ionized radium,
Ra~II~\cite{Fowler1922}. In 1933-34, Rasmussen first measured the
energy levels of singly ionized radium (Ra~II)\cite{Rasmussen1933}
using arc discharge and of neutral radium
(Ra~I)\cite{Rasmussen1934} using 2~mg of radiumchloride
(RaCl$_{2}$) in a discharge lamp and monochromator for determining
the wavelengths. Some of the transition terms of neutral radium
measured by Rasmussen were corrected by
H.~N.~Russell~\cite{Russell1934}. The first compilation of energy
levels for neutral and singly ionized radium was published
by Moore in 1958 in \emph{Atomic Energy Levels}
(AEL)~\cite{Moore1958}. After that, no updates were made to reference data on
neutral and singly ionized Ra. With the
advent of accelerator facilities, modern experimental methods to
produce radium isotopes and to probe the atomic structure, more
information is available on these spectra in
the literature. At this juncture, we realized the need to update
the information available on Ra atom and Ra ion and gathered it,
which might be beneficial for various experiments that are in
progress~\cite{Guest2007,Parker2012,Santra2014,
Portela2014,Parker2015} and for the future
experiments~\cite{Park2014,Kara2014} with different physics
motivations and for various fields of research.

About forty six years after Rasmussen's measurement of energy
levels and wavelengths in Ra ion and Ra atom, furnace-absorption
measurements of energy levels of the $7snp~^{1}P^{0}_{1}$
($n=13-52$) series of neutral radium in the ultraviolet region
were reported by Armstrong et al.~\cite{Armstrong1980,Tomkins1967}. From the experimental results and
applying the multichannel quantum defect theory (MQDT), values for
$7snp$ ($n=9-12$) levels were semi-empirically determined. The reported
uncertainty was 0.006~cm$^{-1}$ for wavenumbers and 0.001~$\AA$
for wavelengths.

The first systematic measurements on a series of neutral
and singly ionized radium isotopes employing the technique of
collinear laser spectroscopy were conducted by the ISOLDE
collaboration at CERN,
Geneva~\cite{Ahmad1983,Arnold1987,Wendt1987,Ahmad1988,Neu1989}.
Radium isotopes were produced by impinging protons of energy
600~MeV on a UC$_{2}$ cloth target. These studies established nuclear spins
and obtained nuclear moments from the measured hyperfine structure
and isotope shifts.

In the last decade, there has been renewed interest in radium
element (radium atom as well as ion) from both theoretical
and experimental physicists. From theoretical calculations, radium
atom is proposed as a promising candidate for searches of a
permanent electric dipole moment in an atomic
system~\cite{Flambaum1999,Dzuba2000,Bieron2007,
Bieron2009,Engel2003,Dobazski2005,Radziute2014}. The reasons are:
large polarizability due to the near degeneracy of the opposite
parity $7s7p$~$^{3}P^{0}_{1}$ and $7s6d$~$^{3}D_{2}$ levels; some
isotopes of radium have nuclear octupole deformation; and high
atomic number ($Z=88$). Singly ionized radium ion is proposed as a
promising candidate for atomic parity violation experiments and as
an atomic clock\cite{Wansbeek2008,Sahoo2009,Versolato2011b}.

In this context, short lived radioactive isotopes were produced at the
TRI$\mu$P facility, University of Groningen, The
Netherlands\cite{Berg2006}. Radium isotopes were produced by the
method of inverse reaction kinematics using $^{206,204}$Pb beams
hitting a rotating carbon target\cite{Traykov2007,Shidling2009}.
Precision laser spectroscopy of radium ions was performed in a
radio frequency quadrupole (RFQ)
trap\cite{Versolato2010,Versolato2011a,Giri2011,Portela2014}.

At the University of Groningen, The Netherlands, with
$^{229}$Th ($10~\mu$Ci) placed in an oven, $^{225}$Ra neutral
atomic beam was produced. Employing this thermal beam of Ra atoms,
the absolute frequency of the strong electric dipole $7s^{2}$~
$^{1}S_{0}$~-~$7s7p$~$^{1}P^{0}_{1}$ transition and weaker
intercombination $7s^{2}$~$^{1}S_{0}$~-~$7s7p$~$^{3}P_{1}$
transition were reported with reference to a femtosecond optical
frequency comb\cite{Santra2014,Santra2012}. The uncertainty
reported was 0.00006~cm$^{-1}$ and 0.00013~cm$^{-1}$, respectively.

At Argonne National Laboratory (ANL), USA, electrochemically
separated neutral radium isotopes, $^{225,226}$Ra placed in an
oven were used for a laser cooling and trapping
experiment\cite{Guest2007}. Further, optical dipole trapping of
$^{226}$Ra\cite{Parker2012} and measurement of lifetimes of
$7s7p$~$^{3}P^{0}_{1}$, $7s6d$~$^{3}D_{1}$ and $7s6d$~$^{1}D_{2}$ states
were reported by the ANL
group\cite{Scielzo2006,Guest2007,Trimble2009}. Very recently,
using cold atoms, the first measurement of the atomic dipole
moment of $^{225}$Ra was reported by the same
group~\cite{Parker2015}.

In this compilation, energy levels, wavelengths, wave numbers,
transition probabilities, lifetimes and other spectral data have
been reviewed. The most accurate information has been assembled from
the available experimental and theoretical data and is presented in Tables~\ref{Raisotopes} - \ref{Raionlifetimes}. The uncertainty
of each value, as given by the original authors, is indicated in
the tables. In the line tables (Tables~\ref{Raatomterms} and \ref{Raionterms}), the wavelengths and their
uncertainties are reported in units of angstroms. For lines
between $2000~\AA$ and $15000~\AA$ air wavelengths and for lines
below $2000~\AA$ vacuum wavelengths are given. The index of
refraction is determined by the three-term formula (Equation 3) of
Peck and Reeder~\cite{Peck1972}. The wave number of each
transition is given in reciprocal centimeters and its uncertainty
is also given wherever applicable. The lower and upper level
columns indicate the classification given for the transition.
 The Ritz wavelengths and their uncertainties
calculated using the level optimization code,
LOPT~\cite{Kramida2011} are also given. The calculated transition
probabilities ($A_{ik}$) are given in units of inverse seconds.
Their accuracy range is also given in a separate column. For
singly ionized Ra uncertainty is given for theoretical transition
rates by Sahoo~\textit{et al.}~\cite{Sahoo2009}
(Table~\ref{Raionterms}).

The energy level tables (Tables~\ref{Raatomenergylevels} and \ref{Raionenergylevels}) contain the configuration,
term, and $J$ values of each energy level, using $LS$ coupling to
describe the configurations. In addition to that, energy
uncertainty, energy level values and their uncertainty from the
LOPT code are also given. For visual clarity, only the first
member of the term has the configuration written out similar to
the notation used by Curry~\cite{Curry2004}. The level value
and its uncertainty are given in reciprocal centimeters. 

Ionization energies are given in units of cm$^{-1}$ and electronvolts (eV). Conversion factor from cm$^{-1}$ to eV was taken from the 2014 CODATA adjustment of fundamental constants~\cite{Codata2014}: 1 cm$^{-1}$ = 1.239 841 9739(76) $\times$ 10$^{-4}$ eV.  The reciprocal centimeter is related to the SI unit for energy, the
joule, in the following way: 1~cm$^{-1}$ is equivalent to $1.986
445824(24)\times10^{-23}$~J~\cite{Codata2014}.

The hyperfine structure constants (HFS) and isotope shifts are
given in Tables~\ref{Raatomhfsc1P1}-\ref{Raatomhfsc3D1}, \ref{Raionhfsc1}-\ref{Raionhfsc3} in units of megahertz (MHz) with the uncertainty in the last
digit given in parentheses following the value. All
uncertainties given in this paper are on the level of one standard
deviation. The hyperfine splitting constant, A, is the magnetic
dipole coefficient, whereas B is the electric quadrupole
coefficient. The references for energy levels and hyperfine
structure splitting and isotope shifts are given to the source of
the energy level value and hyperfine splitting constants,
respectively.

\begin{table*}[htb]
\caption{Isotopes of radium with zero nuclear spin ($I$) and with
non-zero nuclear spin. The half-lives (ns-nanoseconds,
$\mu$s-microseconds, ms-milliseconds, s-seconds, m-minutes, d-days
and y-years) and nuclear spin ($I$) are taken from
Ref.~[\onlinecite{Tuli2011}]. The atomic weights are taken from
Ref.~[\onlinecite{Wang2012}]} \label{Raisotopes}
\begin{ruledtabular}
\begin{tabular}{ccccccc}
  Isotope  & Atomic weight~\cite{Wang2012} & Half-life & Nuclear spin~\cite{Ahmad1983} & Nuclear magnetic & Electric Quadrupole \\
  & u & t$_{1/2}$ & I & moment~\cite{Tuli2011} [$\mu_I/\mu_N$] & moment~\cite{Neu1989} Q$_{s}$ in [b]~\footnotemark[1] \\
  \hline
   $^{235}$Ra    & 235.054970(320)   &             &            &            &          \\
   $^{234}$Ra    & 234.050340(30)    & 30~s        & 0          &            &          \\
   $^{233}$Ra    & 233.047582(17)    & 30~s        &            &            &          \\
   $^{232}$Ra    & 232.043475(10)    & 4.2~m       & 0          &            &          \\
   $^{231}$Ra    & 231.041027(12)    & 104.1~s     & 5/2        &            &          \\
   $^{230}$Ra    & 230.037055(11)    & 93~m        & 0          &            &          \\
   $^{229}$Ra    & 229.034942(16)    & 4.0~m       & 5/2        & 0.5025(27) & 3.09(4) \\
   $^{228}$Ra    & 228.0310707(2.6)  & 5.75~y      & 0          &            &          \\
   $^{227}$Ra    & 227.0291783(2.5)  & 42.2~m      & 3/2        & -0.4038(24)& 1.58(3)   \\
   $^{226}$Ra    & 226.0254103(2.5)  & 1600~y      & 0          &            &          \\
   $^{225}$Ra    & 225.023612(3)     & 14.9~d      & 1/2        & -0.7338(15)&           \\
   $^{224}$Ra    & 224.020212(2.3)   & 3.6319~d    & 0          &            &          \\
   $^{223}$Ra    & 223.0185023(2.7)  & 11.43~d     & 3/2        & 0.2705(19) & 1.254(3)  \\
   $^{222}$Ra    & 222.015375(5)     & 38.0~s      & 0          &            &           \\
   $^{221}$Ra    & 221.013918(5)     & 28~s        & 5/2        & -0.1799(17)& 1.978(7)  \\
   $^{220}$Ra    & 220.011026(9)     & 18~ms       & 0          &            &           \\
   $^{219}$Ra    & 219.010085(9)     & 10~ms       & 7/2        &            &          \\
   $^{218}$Ra    & 218.007141(12)    & 25.2~$\mu$s & 0          &            &         \\
   $^{217}$Ra    & 217.006321(9)     & 1.6~$\mu$s  & 9/2        &            &         \\
   $^{216}$Ra    & 216.003533(9)     & 182~ns      & 0          &            &         \\
   $^{215}$Ra    & 215.002720(8)     & 1.55~ms     & 9/2        &            &         \\
   $^{214}$Ra    & 214.000100(6)     & 2.46~s      & 0          &            &         \\
   $^{213}$Ra    & 213.000384(22)    & 2.73~m      & 1/2        & 0.6133(18) &       \\
   $^{212}$Ra    & 211.999787(12)    & 13~s        & 0          &            &       \\
   $^{211}$Ra    & 211.000893(9)     & 13~s        & 5/2        & 0.8780(38) & 0.48(2)  \\
   $^{210}$Ra    & 210.000494(16)    & 3.7~s       & 0          &            &         \\
   $^{209}$Ra    & 209.001990(50)    & 4.6~s       & 5/2        & 0.865(13)  & 0.40(2)   \\
   $^{208}$Ra    & 208.001841(17)    & 1.3~s       & 0          &            &          \\
   $^{207}$Ra    & 207.003800(60)    & 1.35~s      & 3/2,5/2    &            &          \\
   $^{206}$Ra    & 206.003828(19)    & 0.24~s      & 0          &            &          \\
   $^{205}$Ra    & 205.006270(80)    & 210~ms      & 3/2        &            &          \\
   $^{204}$Ra    & 204.006492(16)    & 57~ms       & 0          &            &          \\
   $^{203}$Ra    & 203.009300(90)    & 31~ms       & 3/2        &            &          \\
   $^{202}$Ra    & 202.009760(26)    & 16~ms       & 0          &            &          \\
   $^{201m}$Ra   & 201.012710(110)   & 1.6~ms      & 13/2       &            &          \\

\end{tabular}
\end{ruledtabular}
\footnotetext[1]{1 barn (b) = $10^{-24}$~cm$^{2}$}
\end{table*}

\section {Ra I}
In Table~\ref{Raisotopes}, radium isotopes, atomic weights,
half-lives, nuclear spins, nuclear magnetic moments and electric
quadrupole moments are given. Up to now 35 radium isotopes have
been discovered excluding the isomers of
$^{203m,205m,207m,213m}$Ra isotopes.

Table~\ref{Raatomenergylevels} presents the energy levels of
$^{226}$Ra. They are taken from Moore 1958
compilation~\cite{Moore1958} and other sources where improved
experimental values are available. Moore's compilation is mainly
based on the measurements of Rasmussen~\cite{Rasmussen1933} and
corrections made to them by Russell~\cite{Russell1934}. In
addition to the compilation of Moore we have added the energy
levels of $^{226}$Ra in the UV region from the absorption
measurements of Armstrong \textit{et al.}~\cite{Armstrong1980}. However, the uncertainty $\pm$0.006~cm$^{-1}$ given
by Armstrong \textit{et al.} for wavenumbers is not correct as the
uncertainty 0.001~$\AA$ in their wavelengths corresponds to the
wavenumber uncertainty of 0.018~cm$^{-1}$. This is brought to the
notice of the authors by Alexander Kramida~\cite{Kramida2015}. The
updated uncertainties of the wavenumbers are given in
Table~\ref{Raatomenergylevels}. We have not included the
semi-empirical values given in Armstrong~\textit{et
al.}~\cite{Armstrong1980} for $7s9p-7s12p$ energy levels. Improved
values of the $7s7p~^{1}P^{0}_{1}$ level from Santra \textit{et
al.,}~\cite{Santra2014} $7s7p~^{3}P^{0}_{1}$ level from Scielzo
\textit{et al.,}~\cite{Scielzo2006} and $7s6d~^{3}D_{1}$ level
from Guest~\textit{et al.,}~\cite{Guest2007} are also given. Furthermore, the measurements of Rasmussen and
Armstrong {\textit et al.} are attributed to the $^{226}$Ra
isotope due to its natural abundance and long half-life compared to the
other Ra isotopes. Also, the high probability that Rasmussen
used $^{226}$Ra is supported by the fact that the knowledge of
isotopes was not available at that time as the first isotope of an
element observed and reported by Urey \textit{et al.} in 1932 was
deuterium~\cite{Urey1932}. Furthermore, the recent measurements
from the Argonne group and Groningen group support this in
addition to the ISOLDE group. The very recent measurement of the
$7s6d$~$^{3}D_{1}$-$7s7p$~$^{1}P^{0}_{1}$ transition in $^{226}$Ra by
the Argonne group~\cite{Guest2007} confirms attribution of Rasmussen's measurements
to the $^{226}$Ra isotope.

The $7p^{2}$ configuration of the $^{1}D_{2}$ term given in
Moore's compilation from Rasmussen measurements was changed to the
$7s7d$ configuration by the ISOLDE group\cite{Wendt1987} based on
the analysis of their experimental results. In
Table~\ref{Raatomenergylevels} of neutral Ra energy level data it
is shown in parentheses. For the $^{1}D_{2}$ term, the
$6d^{2}$ configuration is attributed by Quinet \textit{et
al.}~\cite{Quinet2007}. It is also shown in parentheses. Also, the
$^{3}P_{0,1,2}$ levels assigned to the $7p^{2}$ configuration as
given in Moore's compilation are predicted to belong to the
$6d^{2}$ configuration by Quinet \textit{et
al.}~\cite{Quinet2007}, which is shown in parentheses. For the
$7s8p~^{1}P^{0}_{1}$ level, Armstrong \textit{et
al.}~\cite{Armstrong1980} restored Rasmussen's identification.
Furthermore, Ginges and Dzuba in their recent theoretical
work\cite{Ginges2015} noted that their calculated value of the
$7s8p~^{1}P_{1}$ energy level is very low compared to the values
from Rasmussen's data and Moore's compilation. For example, the
difference from Moore's compilation is 2155~cm$^{-1}$ .
Because of the discrepancies between the observed
spectra and the contradicting interpretations of
configuration terms of some energy levels and deviations from the
theoretical calculations in Ra I more experimental data is
required.

In Table~\ref{Raatomterms} the intensity, observed
wavelengths in air, uncertainty in observed wavelengths, observed
wavenumbers, lower and upper levels, classifications, Ritz
wavelengths obtained with the LOPT code, uncertainties in Ritz
wavelengths and transition probabilities are given. The Ritz wavelengths are not given for lines that alone determine one of the levels involved in the transition. Some of the
observed wavenumbers given in column 4 differ from the wavenumbers given in the original work of Rasmussen~\cite{Rasmussen1934} and Armstrong \textit{et al.}~\cite{Armstrong1980}. The difference from the original values and the present reported values is due to the conversion formulas used by Rasmussen and Armstrong \textit{et al.} for arriving at the wavenumbers from the measured wavelengths. Here we have used the widely accepted formula of Peck and Reeder~\cite{Peck1972} to arrive at the wavenumbers. The first
reference (Column 7) corresponds to the transitions/lines and the
second reference (Column 11) to transition probabilities. We have
given theoretical transition rates for some of the transitions in
Ra atom calculated by Dzuba and Flambaum~\cite{Dzuba2007} and one
value from the experiment~\cite{Scielzo2006}. In column 4 of
Table~\ref{Raatomterms}, we have given wavenumbers for the
transitions from the measurements of
Rasmussen~\cite{Rasmussen1934} and from
Russell~\cite{Russell1934}. Absorption measurements from Armstrong
\textit{et al.}~\cite{Armstrong1980} have no information on the
intensity of the observed lines. Their wavenumbers and air
wavelengths with uncertainties are also given. For the
7s$^{2}$~$^{1}S_{0}$-7s7p~$^{3,1}P^{0}_{1}$ transitions two line
references are given; the first one for the intensity and the
second one for the wavenumber. The wavenumber and the air
wavelength of the 7s6d~$^{3}D_{1}$-7s7p~$^{3}P^{0}_{1}$ transition
measured by the ANL group is also given. The measured value is
6999.84~cm$^{-1}$ with an uncertainty of 0.02~cm$^{-1}$. In
columns 8 and 9, Ritz wavelengths and their uncertainties are
given. These are obtained from the calculations using the LOPT
code of National Institute of Standards and Technology (NIST),
USA.

Theoretically calculated lifetimes and transition rates for some of the energy levels in
neutral Ra have been previously
reported~\cite{Bieron2004,Bieron2007,Dzuba2007}. However, only
three experimental lifetimes of states in neutral Ra atom are
available and they are given in Table~\ref{Raatomlifetimes}.

In Table~\ref{Raatomhfsc1P1}, isotope shifts in frequency for the
$7s^{2}$~$^{1}$S$_{0}$-$7s7p$~$^{1}$P$_{1}$ electric dipole
allowed transition of 19 neutral Ra isotopes and hyperfine
structure constants A and B of the $7s7p$~$^{1}$P$_{1}$ level for
the odd isotopes are given. The isotope shifts are given with
respect to the $^{214}$Ra isotope.

Given in Table~\ref{Raatomhfsc3P1} are the isotope shifts between
eight Ra isotopes for the
$7s^{2}$~$^{1}$S$_{0}$-$7s7p$~$^{3}$P$_{1}$ intercombination
(spin-forbidden) transition and the hyperfine structure constants
$A$ and $B$ of the $7s7p$~$^{3}$P$_{1}$ level for the odd
isotopes. Here again, the isotope shifts are given with respect to
the $^{214}$Ra isotope.

In Table~\ref{Raatomhfsc3P2}, isotope shifts of seven Ra isotopes
for the $7s7p$~$^{3}$P$_{2}$-$7s7d$~$^{3}$D$_{3}$ transition at
644.6~nm and the hyperfine structure constants A and B of the
$7s7p$~$^{3}$P$_{2}$ and $7s7d$~$^{3}$D$_{3}$ levels for the odd
isotopes are given. The isotope shifts are given with respect to
the $^{214}$Ra isotope similar to the other transitions.

In Table~\ref{Raatomhfsc3D1}, the isotope shifts
between $^{226}$Ra and $^{225}$Ra isotopes for the
$7s6d$~$^{3}$D$_{1}$-$7s7p$~$^{1}$P$_{1}$ transition at 1428~nm
and the hyperfine structure constants $A$ and $B$ of the
$7s7p$~$^{1}$P$_{1}$ and $7s6d$~$^{3}$D$_{1}$ levels for the
$^{225}$Ra isotope are given.

\section {Ra II}

The first experimental measurements on singly ionized Ra were
conducted by Rasmussen in 1933 using an arc discharge as the light source. In
total 62 lines were identified, corresponding to 43 energy levels.
The energy levels are given in Table~\ref{Raionenergylevels}. In Table~\ref{Raionenergylevels} additional
information on energy levels and their uncertainties is also
given, which was obtained using the LOPT code~\cite{Kramida2011}.
In column 6 energy level values obtained with the LOPT code are
given. Two uncertainties for the energy level values estimated by
LOPT code are given; uncertainty 1 (Column 7) and uncertainty 2
(Column 8). Uncertainty 1 is the best determined separation from
other energy levels and uncertainty 2 is determined for the
excitation energy. Further, some of the
energy levels given in Moore's compilation~\cite{Moore1958}
are significantly improved by our level optimization, while a few were found to be much less accurate than implied by the number of significant figures given by Moore.

The ionization energy (IE) of Ra II has been
re-determined because of the changes in the energy level values of
Ra II. The series used to determine the IE is $ng~^2G$ with
$n=5-11$, five members of which ($n=5-9$) have been determined
with uncertainties about 0.2~cm$^{-1}$. Using this series, two
values of IE have been obtained; one by fitting a polarization
formula and another by fitting a Ritz-type quantum defect formula
as given in references~\cite{Sansonetti2005,Kramida2013}. The
results of these two determinations are 81842.6(5)~cm$^{-1}$ and
81842.5(5)~cm$^{-1}$ resulting an average of 81842.5(5)~cm$^{-1}$.
One can also use the series $ns~^2S_{1/2}~(n=7-11$),
$nd~^2D_{3/2}~(n=6-11$) and $nd~^2D_{5/2}~(n=6-12$) to determine
IE. However, the IE limit values have larger scatter and
uncertainties. Thus, the recommended IE value is
81842.5(5)~cm$^{-1}$, which agrees with Rasmussen's value of
81842.31~cm$^{-1}$ (as given by Moore) within the uncertainty.

In Table~\ref{Raionterms} intensity, observed
wavelengths of the transitions, uncertainty of the observed
wavelengths, observed wavenumbers, uncertainty of the observed
wavenumbers, the corresponding lower and upper levels, line
reference (Column 8), transition rates, uncertainty of the
transition rates and the reference for the transition rates
(Column 11) are given. Theoretical estimates of energy levels and
dipole matrix elements for singly ionized Ra have been
reported~\cite{Dzuba2001,Quinet2007,Sahoo2009,Pal2009}. Here we
present the transition rates from Sahoo \textit{et
al.}~\cite{Sahoo2009} as the uncertainty was given for the
calculated values. In the intensity (Column 1), 'p' means
'perturbed by'. The corresponding perturbing species are given in
parentheses. In the line reference (Column 8) '6' corresponds to
the reference for the transition/line and the letters; HQ~-~Hilger
Quartz, Q~-~Quartz prism, P~-~Glass prism and G~-~Grating (used in
the red region) are the letter codes for the spectrometer used in
the measurement. The Ritz wavelengths and their uncertainties for
Ra II are given in Table~\ref{Raionritz}. These values were calculated
with the LOPT code~\cite{Kramida2011}. For wavelengths less than
$2000\AA$ vacuum wavelengths are given and for wavelengths greater
than $2000\AA$ air wavelengths are given. The uncertainty of the
given wavelengths mentioned in Ref.~\onlinecite{Rasmussen1933} is
$0.1\AA$ to $0.02\AA$ between 3600 to $7000\AA$ and $0.1\AA$ between 1880 and
$2200\AA$~\cite{Kramida2011}. In the red and near-infrared wavelength regions,
i.e. $7000\AA$ to $10000\AA$ the uncertainty is $0.01\AA$. Some of the
observed wavenumbers given in column 4 differ from the wavenumbers given in the original work of Rasmussen~\cite{Rasmussen1933}. The difference from the original values and the present reported values is due to the conversion formula used by Rasmussen for arriving at the wavenumbers from the measured wavelengths. Here we have used the three-term formula of Peck and Reeder~\cite{Peck1972} to arrive at the wavenumbers.

In Table~\ref{Raisotopesterms}, improved values of wavelengths in
air, wavenumbers and uncertainty in wavenumbers are given for
$^{212,213,214,225,226}$Ra isotopes for the
7s~$^{2}S_{1/2}$-7s7p~$^{2}P^{0}_{1/2,3/2}$ and
6d~$^{2}D_{3/2}$-7s7p~$^{2}P^{0}_{1/2,3/2}$ transitions.
The wavenumbers are derived from the absolute frequency
measurements of the corresponding transitions with a reference to a
femtosecond optical frequency comb. The air wavelengths are
calculated using the three-term formula (Equation 3) from Peck and
Reeder~\cite{Peck1972}.

Table~\ref{Raionhfsc1} gives isotope shifts of 17 Ra isotopes for the
7s~$^{2}S_{1/2}$~-~7p~$^{2}P_{1/2}$ transition and of 8 isotopes
for the 7s~$^{2}S_{1/2}$~-~7p~$^{2}P_{3/2}$ transition, 7
hyperfine structure constants for the 7s~$^{2}S_{1/2}$ and
7p~$^{2}P_{1/2}$ states, while for the 7p~$^{2}P_{3/2}$ state, 3 magnetic dipole constants and 2 electric quadrupole constants measured by the ISOLDE group are given. The uncertainties
are given in parentheses. The isotope shifts are given with
respect to the $^{214}$Ra isotope, which has a zero nuclear spin.

In Table~\ref{Raionhfsc2} isotope shifts of six Ra isotopes for
the 6d~$^{2}D_{3/2}$~-~7p~$^{2}P_{1/2}$ transition of singly
ionized Ra and hyperfine structure constants of the
6d~$^{2}D_{3/2}$ and 7p~$^{2}P_{1/2}$ states measured at the
University of Groningen are given.

Table~\ref{Raionhfsc3} consists of isotope shifts of the $^{212,
213}$Ra isotopes with respect to $^{214}$Ra for the
6d~$^{2}D_{3/2}$~-~7p~$^{2}P_{3/2}$ transition and hyperfine
structure of the 6d~$^{2}D_{3/2}$ state. There is only one
experimental measurement of lifetime of a state in singly ionized
Ra. It is for the 6d~$^{2}D_{5/2}$ state in $^{212}$Ra with the
lifetime of 232(4)~ms (Table~\ref{Raionlifetimes}). Although
theoretical calculations are available for the lifetimes of
different states in the
literature~\cite{Dzuba2001,Sahoo2007,Sahoo2009,Pal2009}, we have
not included them in this compilation.

\section{Spectroscopic Data}

Ground state and configuration:

Ra I~-~$1s^2~2s^{2}2p^{6}~3s^{2}3p^{6}3d^{10}~4s^{2}4p^{6}4d^{10}4f^{14}\\
~5s^{2}5p^{6}5d^{10}~6s^{2}6p^{6}~7s^{2}$~$^{1}S_{0}$\\

Ra II~-~$1s^2~2s^{2}2p^{6}~3s^{2}3p^{6}3d^{10}~4s^{2}4p^{6}4d^{10}4f^{14}\\
~5s^{2}5p^{6}5d^{10}~6s^{2}6p^{6}~7s$~$^{2}S_{1/2}$\\

Ionization energies:

Ra I~-~42573.36(2)~cm$^{-1}$;~
5.2784239(25)~eV~\cite{Armstrong1980}

Ra II~-~81842.5(5)~cm$^{-1}$;~
10.14718(6)~eV

\begin{longtable*}{*{8}{>{\centering\arraybackslash}p{2.1cm}}}
\caption{Energy levels of $^{226}$Ra I.}\\
 \hline \hline
\multicolumn{1}{c}{} & \multicolumn{1}{c}{} & \multicolumn{1}{c}{}
& \multicolumn{1}{c}{} &\multicolumn{1}{c}{} &\multicolumn{1}{c}{}
& \multicolumn{1}{c}{} & \multicolumn{1}{c}{}\\
\endfirsthead

 \multicolumn{8}{c}
{{ \tablename\ \thetable{} Energy levels, $^{226}$Ra I -- continued}} \\
\hline \hline \multicolumn{1}{c}{} & \multicolumn{1}{c}{} &
\multicolumn{1}{c}{} & \multicolumn{1}{c}{Level} &
\multicolumn{1}{c}{Uncertainty} & \multicolumn{1}{c}{Level LOPT} &\multicolumn{1}{c}{Uncertainty} & \multicolumn{1}{c}{Level} \\
\multicolumn{1}{c}{Configuration} & \multicolumn{1}{c}{Term} &
\multicolumn{1}{c}{J} & \multicolumn{1}{c}{(cm$^{-1}$)} &
\multicolumn{1}{c}{(cm$^{-1}$)} & \multicolumn{1}{c}{(cm$^{-1}$)} & \multicolumn{1}{c}{LOPT (cm$^{-1}$)} & \multicolumn{1}{c}{reference} \\
\hline
\endhead
               &      &    & Level       & Uncertainty & Level LOPT  & Uncertainty      & Level   \\
 Configuration & Term &  J & (cm$^{-1}$) & (cm$^{-1}$) & (cm$^{-1}$) & LOPT (cm$^{-1}$) & reference  \\
\hline
7s$^{2}$      & $^{1}$S  & 0 & 0.0000       &       & 0.0000   & 0.0    &      \\

7s7p     & $^{3}$P$^{0}$ & 0 & 13078.44  &          & 13078.40 & 0.04 & \onlinecite{Moore1958}  \\
         & $^{3}$P$^{0}$ & 1 & 13999.357 & 0.001    & 13999.3569 & 0.0010 & \onlinecite{Scielzo2006} \\
         & $^{3}$P$^{0}$ & 2 & 16688.54  &          & 16688.54 & 0.04 & \onlinecite{Moore1958} \\

7s6d     & $^{3}$D       & 1 & 13715.85 &           & 13715.779 & 0.020 & \onlinecite{Moore1958} \\
         & $^{3}$D       & 2 & 13993.97 &           & 13993.94  & 0.04  & \onlinecite{Moore1958} \\
\end{longtable*}


\begin{longtable*}{*{13}{>{\centering\arraybackslash}p{1.55cm}}}
\caption{Wavelengths, Wave numbers, classification and transition
probabilities in neutral radium ($^{226}$Ra), Ra I}\\
 \hline \hline
\multicolumn{1}{c}{} & \multicolumn{1}{c}{} & \multicolumn{1}{c}{}
& \multicolumn{1}{c}{} & \multicolumn{1}{c}{} &
\multicolumn{1}{c}{} & \multicolumn{1}{c}{} & \multicolumn{1}{c}{}
& \multicolumn{1}{c}{} & \multicolumn{1}{c}{} & \multicolumn{1}{c}{}& \multicolumn{1}{c}{}\\
\endfirsthead

 \multicolumn{13}{c}
{{\tablename\ \thetable{} Wavelengths, Wave numbers,
classification and transition
probabilities in $^{226}$Ra I -- continued}} \\
\hline \hline \multicolumn{1}{c}{} & \multicolumn{1}{c}{Observed}
& \multicolumn{1}{c}{Unc.} & \multicolumn{1}{c}{Observed} &
\multicolumn{1}{c}{Lower} &
\multicolumn{1}{c}{Upper} & \multicolumn{1}{c}{Line} & Ritz & Unc. Ritz & \multicolumn{1}{c}{} & \multicolumn{1}{c}{} \\
\multicolumn{1}{c}{Intensity} &
\multicolumn{1}{c}{$\lambda_{air}/\AA$} & \multicolumn{1}{c}{Obs.
$\lambda_{air}/\AA$} & \multicolumn{1}{c}{$\sigma$/cm$^{-1}$} &
\multicolumn{1}{c}{level} &
\multicolumn{1}{c}{level} & \multicolumn{1}{c}{reference(s)} & $\lambda/\AA$ & $\lambda/\AA$ & \multicolumn{1}{c}{A$_{ik}$/s$^{-1}$} & \multicolumn{1}{c}{Reference} \\
\hline
\endhead
           & Observed                & Unc. Obs.               & Obs.               & Lower  & Upper & Line         & Ritz              & Unc. Ritz   & \\
Intensity  & $\lambda$$_{air}$/$\AA$ & $\lambda$$_{air}$/$\AA$ & $\sigma$/cm$^{-1}$ & level  & level & reference(s) &  $\lambda$/$\AA$  & $\lambda$/$\AA$ & A$_{ik}$/s$^{-1}$ & Reference \\
 \hline
& 2350.8675 & 0.0010 & 42524.473 & $7s^{2}$~$^{1}$S$_{0}$ & $7s52p$~$^{1}$P$^{0}_{1}$ & \onlinecite{Armstrong1980} &  &  & \\
& 2350.9808 & 0.0010 & 42522.423 & $7s^{2}$~$^{1}$S$_{0}$ & $7s51p$~$^{1}$P$^{0}_{1}$ & \onlinecite{Armstrong1980} &  &  & \\
& 2351.1079 & 0.0010 & 42520.125 & $7s^{2}$~$^{1}$S$_{0}$ & $7s50p$~$^{1}$P$^{0}_{1}$ & \onlinecite{Armstrong1980} &  &  & \\
& 2351.2363 & 0.0010 & 42517.803 & $7s^{2}$~$^{1}$S$_{0}$ & $7s49p$~$^{1}$P$^{0}_{1}$ & \onlinecite{Armstrong1980} &  &  & \\
& 2351.3841 & 0.0010 & 42515.131 & $7s^{2}$~$^{1}$S$_{0}$ & $7s48p$~$^{1}$P$^{0}_{1}$ & \onlinecite{Armstrong1980} &  &  & \\
& 2351.5429 & 0.0010 & 42512.260 & $7s^{2}$~$^{1}$S$_{0}$ & $7s47p$~$^{1}$P$^{0}_{1}$ & \onlinecite{Armstrong1980} &  &  & \\
\end{longtable*}

\begin{table*}[!h]
\caption{Lifetimes of energy levels - $^{226}$Ra I}
\label{Raatomlifetimes}
\begin{ruledtabular}
\begin{tabular}{ccccccccc}
Level  & 7s7p~$^{3}$P$^{0}_{1}$ &
7s6d~$^{1}$D$_{2}$ &  7s6d~$^{3}$D$_{1}$ \\
 \hline
Lifetime &  422(20)~ns & 385(45)~$\mu$s & 510(60)~$\mu$s \\

Reference & \onlinecite{Scielzo2006} & \onlinecite{Trimble2009} & \onlinecite{Guest2007} \\

\end{tabular}
\end{ruledtabular}
\end{table*}

\begin{table*}[!htbp]
\caption{Isotope shifts and hyperfine structure constants A and B
of 7s$^{2}$~$^{1}$S$_{0}$ and 7s7p~$^{1}$P$^{0}_{1}$ levels in Ra
I} \label{Raatomhfsc1P1}
\begin{ruledtabular}
\begin{tabular}{ccccccccc}
      &  7s$^{2}$~$^{1}$S$_{0}$   &   &  7s7p~$^{1}$P$^{0}_{1}$ & &     \\
      \cline{2-2} \cline{3-5}  \\
Atomic      &       &   Isotopic frequency  & HFS &  HFS &  \\
mass   & & shift (MHz)  & A (MHz) & B (MHz)    & Reference \\
 \hline

208          &    & 8485(8)     &               &             & \onlinecite{Wendt1987}   \\
209          &    & 8313(8)     & -656.5(1.2)   & 135.0(5.0)  & \onlinecite{Wendt1987}  \\
210          &    & 6003(6)     &               &             & \onlinecite{Wendt1987}  \\
211          &    & 5522(4)     & -668.4(0.9)   & 162.0(4.0)  & \onlinecite{Wendt1987} \\
\end{tabular}
\end{ruledtabular}
\end{table*}

\begin{table*}[!htbp]
\caption{Isotope shifts and hyperfine structure constants A and B
of 7s$^{2}$~$^{1}$S$_{0}$ and 7s7p~$^{3}$P$^{0}_{1}$ levels in Ra
I} \label{Raatomhfsc3P1}
\begin{ruledtabular}
\begin{tabular}{ccccccccc}
      &  7s$^{2}$~$^{1}$S$_{0}$   &   &  7s7p~$^{3}$P$^{0}_{1}$ & &     \\
      \cline{2-2}  \cline{3-5} \\
Atomic      &       &   Isotopic frequency  & HFS &  HFS &  \\
mass   &    & shift (MHz)  & A (MHz) & B (MHz)  & Reference \\
 \hline

212          &     & 3303(5)     &               &             & \onlinecite{Wendt1987}  \\
214          &  0  &  0          &               &             & \onlinecite{Wendt1987}  \\
221          &     &             & -475.8(1.4)   & -747.6(2.6) & \onlinecite{Wendt1987}  \\
222          &     & -29253(9)   &               &             & \onlinecite{Wendt1987} \\
\end{tabular}
\end{ruledtabular}
\end{table*}


\begin{table*}[!htbp]
\caption{Isotope shifts and hyperfine structure constants A and B
of the 7s7p~$^{3}$P$^{0}_{2}$ level and the 7s7d~$^{3}$D$_{3}$
level in Ra I} \label{Raatomhfsc3P2}
\begin{ruledtabular}
\begin{tabular}{ccccccccc}
      &  7s7p~$^{3}$P$^{0}_{2}$   &   & & 7s7d~$^{3}$D$_{3}$ & &     \\
      \cline{2-3} \cline{4-6}  \\
Atomic      &  HFS          & HFS     &   Isotopic             & HFS     &  HFS       &  \\
mass  & A(MHz) & B (MHz) & frequency shift (MHz)  & A (MHz) & B (MHz)    & Reference \\
 \hline

212   &           &               & -887(4)     &               &             & \onlinecite{Wendt1987} \\
214   &           &               &   0         &               &             & \onlinecite{Wendt1987}  \\
221   & 278.6(0.9)& 1103.8(3.9)   & 7001(7)     & 213.3(0.6)    & 117.8(4.4)  & \onlinecite{Wendt1987}  \\
222   &           &               & 7781(4)     &               &             & \onlinecite{Wendt1987} \\
\end{tabular}
\end{ruledtabular}
\end{table*}


\begin{table*}[!htbp]
\caption{Isotope shifts and hyperfine structure constants A and B
of the 7s7p~$^{1}$P$^{0}_{1}$ level and the 7s6d~$^{3}$D$_{1}$
level in Ra I} \label{Raatomhfsc3D1}
\begin{ruledtabular}
\begin{tabular}{ccccccccc}
      &  7s7p~$^{1}$P$^{0}_{1}$   &   & & 7s6d~$^{3}$D$_{1}$ & &     \\
      \cline{2-3} \cline{4-6}  \\
Atomic  &  HFS    & HFS     &   Isotopic             & HFS      & HFS       &  \\
mass    & A(MHz)  & B (MHz) & frequency shift (MHz)  & A (MHz)  & B (MHz)   & Reference \\
 \hline

225   &   2797.3 (1.5)  &               & 540.2 (2.0)          &   4687.7 (1.5)  &    & \onlinecite{Guest2007}  \\
226   &                 &               & 0                    &                 &    & \onlinecite{Guest2007} \\

\end{tabular}
\end{ruledtabular}
\end{table*}

\clearpage

\begin{table*}[!h]
\caption{Energy levels of $^{226}$Ra II.}
\label{Raionenergylevels}
\begin{ruledtabular}
\begin{tabular}{cccccccccc}
 &  &  & Level & Uncertainty & Level LOPT & Unc.~1 LOPT & Unc.~2 LOPT & Level  \\
Configuration & Term & J &(cm$^{-1}$) & (cm$^{-1}$) & (cm$^{-1}$) & (cm$^{-1}$) & (cm$^{-1}$) & reference  \\
 \hline
7s   & $^{2}$S       & 1/2 & 0.000    &  & 0.000      & 0.0007 & 0 &    \\

6d   & $^{2}$D       & 3/2 & 12084.2721 &  & 12084.2721 & 0.0006 & 0.0010 & \onlinecite{Moore1958} \\
     & $^{2}$D       & 5/2 & 13742.99 &  & 13742.99   & 0.03   & 0.03   & \onlinecite{Moore1958} \\

7p   & $^{2}$P$^{0}$ & 1/2 & 21351.3259 & 0.0008 & 21351.3259  & 0.0005 & 0.0008 & \onlinecite{Moore1958}, \onlinecite{Portela2014} \\
     & $^{2}$P$^{0}$ & 3/2 & 26208.8538 & 0.0021 & 26208.8538  & 0.0021 & 0.0021 & \onlinecite{Moore1958}, \onlinecite{Portela2014} \\

8s   & $^{2}$S       & 1/2 & 43405.05 & & 43405.05 & 0.05 & 0.05 & \onlinecite{Moore1958}  \\
\end{tabular}
\end{ruledtabular}
\end{table*}

\begin{longtable*}{*{11}{>{\centering\arraybackslash}p{1.71cm}}}
\caption{Wavelengths, Wave numbers, classification and transition
probabilities for observed lines in radium ion, $^{226}$Ra II} \\
 \hline \hline
\multicolumn{1}{c}{} & \multicolumn{1}{c}{} & \multicolumn{1}{c}{}
& \multicolumn{1}{c}{} & \multicolumn{1}{c}{} &
\multicolumn{1}{c}{} & \multicolumn{1}{c}{} &
\multicolumn{1}{c}{} & \multicolumn{1}{c}{} & \multicolumn{1}{c}{}& \multicolumn{1}{c}{}\\
\endfirsthead

 \multicolumn{11}{c}
{{\tablename\ \thetable{} Wavelengths, Wave numbers,
classification and
transition probabilities in radium ion, $^{226}$Ra II-- continued}} \\
\hline \hline \multicolumn{1}{c}{} & \multicolumn{1}{c}{Obs.} &
\multicolumn{1}{c}{Unc. Obs.} & \multicolumn{1}{c}{Obs.}   &
\multicolumn{1}{c}{Unc. Obs.} & \multicolumn{1}{c}{Lower} &
\multicolumn{1}{c}{Upper} & \multicolumn{1}{c}{Line} &
\multicolumn{1}{c}{} & \multicolumn{1}{c}{} & \multicolumn{1}{c}{} \\
\multicolumn{1}{c}{Intensity} &
\multicolumn{1}{c}{$\lambda_{air}/\AA$} & \multicolumn{1}{c}{
$\lambda_{air}/\AA$ } &\multicolumn{1}{c}{$\sigma$/cm$^{-1}$}  &
\multicolumn{1}{c}{$\sigma$/cm$^{-1}$} &
\multicolumn{1}{c}{level}&
\multicolumn{1}{c}{level} & \multicolumn{1}{c}{ref.} &\multicolumn{1}{c}{A$_{ik}$/s$^{-1}$} & \multicolumn{1}{c}{Acc.} & \multicolumn{1}{c}{Ref.} \\
\hline

\endhead

 &  Obs. & Unc. Obs. & Obs. & Unc. Obs. & Lower & Upper & Line &  &  &  \\
Intensity~\footnotemark[2] & $\lambda$/$\AA$ & $\lambda$/$\AA$ & $\sigma$/cm$^{-1}$ & $\sigma$/cm$^{-1}$ & level & level & ref.~\footnotemark[3] & $A_{ik}$/s$^{-1}$ & Acc.~\footnotemark[4] & Ref. \\
 \hline

5   & 1888.70  & 0.20 & 52946 & & 6d $^{2}$D$_{5/2}$ & 7f $^{2}$F$^{0}$$_{7/2}$  & \onlinecite{Rasmussen1933}HQ & & & \\
8   & 1908.70  & 0.20 & 52392 & & 7s $^{2}$S$_{1/2}$ & 8p $^{2}$P$^{0}$$_{3/2}$  & \onlinecite{Rasmussen1933}HQ & $1.75\times 10^{7}$ & B & \onlinecite{Sahoo2009} \\
4   & 1972.60  & 0.20 & 50695 & & 7p $^{2}$P$^{0}$$_{1/2}$ & 10d $^{2}$D$_{3/2}$ & \onlinecite{Rasmussen1933}HQ & & & \\
2   & 1976.00  & 0.20 & 50607 & & 7s $^{2}$S$_{1/2}$ & 8p $^{2}$P$^{0}$$_{1/2}$  & \onlinecite{Rasmussen1933}HQ & &  & \\
2   & 2006.40  & 0.20 & 49824 & & 7p $^{2}$P$^{0}$$_{1/2}$ & 11s $^{2}$S$_{1/2}$ & \onlinecite{Rasmussen1933}HQ & & & \\
3   & 2012.75  & 0.10 & 49667 & & 6d $^{2}$D$_{5/2}$ & 9p $^{2}$P$^{0}$$_{3/2}$  & \onlinecite{Rasmussen1933}HQ & & & \\
\end{longtable*}

\begin{table*}[!htbp]
\caption{Ritz wavelengths and uncertainty of Ritz wavelengths in
$^{226}$Ra II} \label{Raionritz}
\begin{ruledtabular}
\begin{tabular}{ccccc}

  Lower & Upper & Ritz & Uncertainty \\
level & level & $\lambda$/$\AA$ & Ritz $\lambda$/$\AA$  \\
 \hline
 6d $^{2}$D$_{5/2}$ & 7f $^{2}$F$^{0}$$_{7/2}$  & 1888.639 & 0.015  \\
 7s $^{2}$S$_{1/2}$ & 8p $^{2}$P$^{0}$$_{3/2}$  & 1908.688 & 0.007  \\
 7p $^{2}$P$^{0}$$_{1/2}$ & 10d $^{2}$D$_{3/2}$ & 1972.698 & 0.012  \\
 7s $^{2}$S$_{1/2}$ & 8p $^{2}$P$^{0}$$_{1/2}$  & 1976.039 & 0.009  \\
 7p $^{2}$P$^{0}$$_{1/2}$ & 11s $^{2}$S$_{1/2}$ & 2006.514 & 0.011   \\
 6d $^{2}$D$_{5/2}$ & 9p $^{2}$P$^{0}$$_{3/2}$  & 2012.741 & 0.008  \\
\end{tabular}
\end{ruledtabular}
\end{table*}

\begin{table*}[!htbp]
\caption{Wavelengths, wavenumbers and terms of isotopes of Ra II}
\label{Raisotopesterms}
\begin{ruledtabular}
\begin{tabular}{ccccccccc}
Atomic &  &  & Uncertainty & Lower &
Upper &  &   \\
Mass & $\lambda_{air}$/$\AA$ & $\sigma$/cm$^{-1}$ & (cm$^{-1}$) & level & level & Reference \\
 \hline

212 & 3814.0962 & 26211.0916 & 0.0021 & 7s $^{2}$S$_{1/2}$ & 7p $^{2}$P$^{0}$$_{3/2}$   & \onlinecite{Portela2014} \\
    & 4681.78273  & 21353.4085 & 0.0005 & 7s $^{2}$S$_{1/2}$ & 7p $^{2}$P$^{0}$$_{1/2}$ & \onlinecite{Portela2014} \\
    & 7078.0714 & 14124.2481  & 0.0018 & 6d $^{2}$D$_{3/2}$ & 7p $^{2}$P$^{0}$$_{3/2}$    & \onlinecite{Portela2014} \\
    & 10788.530 & 9266.5650  & 0.0004 & 6d $^{2}$D$_{3/2}$ & 7p $^{2}$P$^{0}$$_{1/2}$  & \onlinecite{Portela2014} \\

213 & 4681.79396  & 21353.3573 & 0.0004 & 7s $^{2}$S$_{1/2}$ & 7p $^{2}$P$^{0}$$_{1/2}$ & \onlinecite{Portela2014} \\
    & 7078.0822 (F=2-F=1)  &  14124.2264 & 0.0017 & 6d $^{2}$D$_{3/2}$ & 7p $^{2}$P$^{0}$$_{3/2}$ & \onlinecite{Portela2014} \\
    & 10788.5176  & 9266.5756 & 0.0005 & 6d $^{2}$D$_{3/2}$ & 7p $^{2}$P$^{0}$$_{1/2}$  &\onlinecite{Portela2014} \\

214 & 3814.12014  & 26210.9273 & 0.0017 & 7s $^{2}$S$_{1/2}$ & 7p $^{2}$P$^{0}$$_{3/2}$   & \onlinecite{Portela2014} \\
    & 4681.81626   & 21353.2556 & 0.0005 & 7s $^{2}$S$_{1/2}$ & 7p $^{2}$P$^{0}$$_{1/2}$ & \onlinecite{Portela2014} \\
    & 7078.0599  & 14124.2709  & 0.0014 & 6d $^{2}$D$_{3/2}$ & 7p $^{2}$P$^{0}$$_{3/2}$    & \onlinecite{Portela2014} \\
    & 10788.4902  & 9266.5992  & 0.0004 & 6d $^{2}$D$_{3/2}$ & 7p $^{2}$P$^{0}$$_{1/2}$  & \onlinecite{Portela2014} \\

225    & 4682.21642  & 21351.4307 &0.0008 & 7s $^{2}$S$_{1/2}$ & 7p $^{2}$P$^{0}$$_{1/2}$ & \onlinecite{Portela2014} \\
       & 10787.9892  & 9267.0295 & 0.0006 & 6d $^{2}$D$_{3/2}$ & 7p $^{2}$P$^{0}$$_{1/2}$ & \onlinecite{Portela2014} \\

226    & 10787.9609  & 9267.0538 & 0.0006 & 6d $^{2}$D$_{3/2}$ & 7p $^{2}$P$^{0}$$_{1/2}$ & \onlinecite{Portela2014} \\

\end{tabular}
\end{ruledtabular}
\end{table*}


\begin{table*}[!htbp]
\begin{center}
\caption{Isotope shifts and hyperfine constants of
7s~$^{2}$S$_{1/2}$,  7p~$^{2}$P$^{0}_{1/2}$, and
7p~$^{2}$P$^{0}_{3/2}$ levels in Ra II} \label{Raionhfsc1}
\begin{ruledtabular}
\begin{tabular}{ccccccccc}
        &  7s~$^{2}$S$_{1/2}$     &  7p~$^{2}$P$^{0}_{1/2}$ & & & 7p~$^{2}$P$^{0}_{3/2}$    \\
\cline{2-2}  \cline{3-4}  \cline{5-7} \\
 Atomic           & HFS     & Isotopic frequency               & HFS       & Isotopic frequency           & HFS      & HFS  \\
 mass & A (MHz) & shift (MHz)  &  A (MHz)  & shift (MHz)\footnotemark[1]        & A (MHz)\footnotemark[1]  & B (MHz)\footnotemark[1] & Reference(s)\\
 \hline
208 &             & 11950(8)     &              &                             &            &          & \onlinecite{Wendt1987} \\
209 &             &              &              &                             &            &          & \onlinecite{Wendt1987} \\
210 &             & 8449(6)      &              &                             &            &          & \onlinecite{Wendt1987} \\
211 & 6624.8(1.0) & 7770(4)      & 1299.7(0.8)  &                             &            &          & \onlinecite{Wendt1987} \\
\end{tabular}
\end{ruledtabular}
\end{center}
\footnotetext[1]{From reference~\onlinecite{Neu1989}}
\end{table*}

\begin{table*}[!htbp]
\begin{center}
\caption{Isotope shifts and hyperfine structure constants A and B
of the 6d~$^{2}$D$_{3/2}$ and 7p~$^{2}$P$^{0}_{1/2}$ levels in Ra
II} \label{Raionhfsc2}
\begin{ruledtabular}
\begin{tabular}{ccccccccc}
      &  6d~$^{2}$D$_{3/2}$   &   &  7p~$^{2}$P$^{0}_{1/2}$ & & &     \\
      \cline{2-3} \cline{4-5}   \\
 Atomic           & HFS    & HFS     &   Isotopic             & HFS            &  \\
 mass & A (MHz) & B (MHz) & frequency shift (MHz)  & A (MHz)        & Reference(s) \\
 \hline

209    & 148(10)\footnotemark[1] & 104(38)\footnotemark[1] & 2645(56)\footnotemark[2] & 1276(20)\footnotemark[4] & \onlinecite{Versolato2010}, \onlinecite{Versolato2011a}, \onlinecite{Giri2011} \\
210    &                             &                             & 1884(16)                   &                               & \onlinecite{Giri2011}  \\
211    & 151(2)\footnotemark[1]  & 103(6)\footnotemark[1]  & 1755(14)\footnotemark[2] & 1299.7(0.8)\footnotemark[3]  &  \onlinecite{Versolato2011a}, \onlinecite{Giri2011}, \onlinecite{Wendt1987} \\
212    &                             &                             & 1025(12) &                               & \onlinecite{Giri2011}  \\
213    & 528(5)\footnotemark[4]  &                             & 707(14)\footnotemark[2]  & 4525(5)\footnotemark[3]      &   \onlinecite{Versolato2010}, \onlinecite{Giri2011}, \onlinecite{Wendt1987}\\
214    &                             &                             &  0                       &                               &   \\

\end{tabular}
\end{ruledtabular}
\end{center}
\footnotetext[1]{Reference~\onlinecite{Versolato2011a}}\footnotetext[2]{Reference~\onlinecite{Giri2011}}\footnotetext[3]{Reference~\onlinecite{Wendt1987}}
\footnotetext[4]{Reference~\onlinecite{Versolato2010}}
\end{table*}


\begin{table*}[!htbp]
\caption{Isotope shifts and hyperfine structure constants A and B
of the 7p~$^{2}$P$^{0}_{3/2}$ and 6d~$^{2}$D$_{3/2}$ levels
in Ra II} \label{Raionhfsc3}
\begin{ruledtabular}
\begin{tabular}{ccccccccc}
      &  6d~$^{2}$D$_{3/2}$   &   &  7p~$^{2}$P$^{0}_{3/2}$ & &     \\
      \cline{2-2} \cline{3-5}  \\
Atomic          & HFS    &   Isotopic             & HFS     & HFS       &  \\
 mass & A(MHz) & frequency shift (MHz)  & A (MHz) & B (MHz)    & Reference(s) \\
 \hline

212    & 528(5) & 701(50)                 &  &  & \onlinecite{Versolato2010}, \onlinecite{Giri2011} \\
213    &        & 453(34)                 &  &  & \onlinecite{Versolato2010} \\
214    &        &  0                      &  &  &  \\

\end{tabular}
\end{ruledtabular}
\end{table*}


\begin{table*}[!htbp]
\caption{Lifetimes of energy levels in $^{212}$Ra II}
\label{Raionlifetimes}
\begin{ruledtabular}
\begin{tabular}{ccccccccc}
Level     & 6d~$^{2}$D$_{5/2}$   \\
 \hline
Lifetime  & 232(4)~ms  \\

Reference &  \onlinecite{Versolato2010}  \\

\end{tabular}
\end{ruledtabular}
\end{table*}

\begin{acknowledgments}
The authors gratefully acknowledge Alexander~Kramida,
Atomic Spectroscopy Group, National Institute of Standards and
Technology (NIST), USA for generously providing his calculations
of ionization energy, LOPT code and
other information. One of the authors U.\ D.\ acknowledges kind
help from Arun K.\ Thazathveetil, Department of Chemistry,
North-Western University, Evanston, USA, S. Knoop, Vrije
Universiteit, Amsterdam, The Netherlands, Lotje Wansbeek, The
Netherlands and B. Santra, Technische Universit\"{a}t
Kaiserslautern, Germany.
\end{acknowledgments}

\bibliography{JPhyschem-ref}

\end{document}